\documentclass[12pt]{article}
\usepackage{graphicx}
\usepackage{url}
\usepackage{hyperref}
\usepackage{amsmath}
\usepackage{amssymb}
\usepackage{geometry}
\usepackage[hang]{footmisc} 
\usepackage{parskip} 
\title{On the Preservation of Africa's Cultural Heritage in the Age of Artificial Intelligence}
\author{Mohamed El Louadi \\
University of Tunis \\
Institut Supérieur de Gestion \\
41 rue de la Liberté – Cité Bouchoucha \\
2000 Le Bardo -- Tunisia \\
\href{mailto:mohamed.louadi@isg.rnu.tn}{mohamed.louadi@isg.rnu.tn}}

\date{}

\begin{document}

\maketitle

\begin{abstract}
In this paper, we delve into the historical evolution of data as a fundamental element in communication and knowledge transmission. The paper traces the stages of knowledge dissemination from oral traditions to the digital era, highlighting the significance of languages and cultural diversity in this progression. It also explores the impact of digital technologies on memory, communication, and cultural preservation, emphasizing the need for promoting a culture of the digital (rather than a digital culture) in Africa and beyond. Additionally, it discusses the challenges and opportunities presented by data biases in AI development, underscoring the importance of creating diverse datasets for equitable representation. We advocate for investing in data as a crucial raw material for fostering digital literacy, economic development, and, above all, cultural preservation in the digital age.

\end{abstract}

\section{Introduction}
In 2017, the special issue of \textit{The Economist} announced on its cover that the most valuable resource was now data. This echoed the phrase ''data is the new oil,'' coined in 2006 by data scientist Clive Humby. 

In 2018, James Bridle clarified that data is not the new oil but the new nuclear energy in that it is unlimited and in its propensity to harm \cite{Bridle2018}. 

Since the 1960s, the following terms have followed one another at a frantic pace: database, data warehouse, data mining, databank, datamart, data store, big data, data lake, dataset, all Anglo-Saxon-sounding terms with sometimes more or less successful attempts to translate them into other languages. \\

But since the dawn of time, data has been the basis of communication and the transmission of knowledge between humans, which evolved in four stages: oral, written, printed, and digital.

\section{The Stages of Knowledge Transmission}
North of Vancouver, there lives a tribe that believes every time a person is killed, a story is lost. This belief is shared by many cultures and civilizations, where the past was transmitted orally, around a campfire, or deep in a cave, when our ancestors sought refuge from the elements. These stories were memorized and passed down from one generation to the next. 

To this day, oral tradition continues to be a fundamental aspect of the traditional culture not only of the Squamish but of many other cultures and many other geographies. By 2010, the Squamish language had all but disappeared as it was spoken by only ten of the remaining 3,900 individuals. One day soon, the story of more than a millennium of this tribe, and others, will no longer be told orally, or even in its original language, with the loss of meaning and subtleties that one can guess.

\subsection{Languages, a Vector of Oral Transmission}
Are languages disappearing? 

While just 1000 years ago there were nearly 9000 languages spoken in the world \cite{WorldDataInfo}, there are only between 6500 \cite{WorldDataInfo} and 7168 \cite{CIAGov2024} left, including 2140 in Africa \cite{CIAGov2024}. Experts predict that by 2050 there will be only about 4500 languages, 3000 in 2100, and 100 by the onset of the 23rd century \cite{CIAGov2024}. 

While stories were transmitted orally through language, humans had an impressive memory compared to ours. It is believed that our ancestors could memorize and recite a poem simply by reading it once. Troubadours as well as Arab poets competed with each other, and memory was appreciated, if not adored. 

At some point in the history of mankind, the written word appeared, in all likelihood, in Ancient Mesopotamia.

\section{The Written Word}
It is said that Socrates despised writing, thinking that it encouraged laziness and diminished memory. It is well known that this reasoning opposed him to Plato who, in his play \textit{Phaedrus}, depicts Socrates as deploring the development of writing\footnote{See pages xiv and xv of Phaedrus by Plato (1978). W.C. Helmbold and W.G. Rabinovitz, The Library of Liberal Arts, Indianapolis: Bobbs-Merrill, \url{(https://archive.org/details/phaedrus0000plat_z9y2/page/n15/mode/2up?q=writing)}, January 17, 2023.}\footnote{Plato had mixed opinions about writing. In his dialogue Phaedrus, he argues that writing is inferior to oral tradition because it is a static and immutable representation of knowledge. Oral tradition allows for the possibility of interactive dialogue. However, in other dialogues, such as the Theaetetus and the Sophist, Plato concedes that writing is useful as a means of preserving and transmitting knowledge.}. 

Diogenes also considered writing to be inferior to speech, which allows for more authentic and immediate communication. According to him, writing freezes thought and allows individuals to hide their true thoughts behind written words and manipulate the truth. Paradoxically, Diogenes did not deprive himself of writings, but none of his texts survived him, some because he burned them. What we know today of Diogenes' writings has been passed down to us through fragments in oral testimony. 

Thus we divested ourselves of enormous memory capacities when we began to write (or draw) and, later, to record knowledge. 

Thus, if languages vanish, human memory depletes as well.

Humans have become accustomed to keeping in their memory only what is not accessible elsewhere. They then began to offload their memory into devices. This is how we forgot the phone numbers of our loved ones when we bought mobile phones; GPS systems may have diminished our ability to read maps and spell checkers or calculators may have contributed to the emergence of a generation that is no longer comfortable with the rules of spelling, grammar or mental arithmetic. We only internalize in our bodies what cannot be found elsewhere. Soon technologies will host everything that is supposed to be in our heads: memory, reasoning, imagination, etc. 

Then came the printing press.

\section{The Printing Press}
The printing press was invented by the Chinese and popularized a thousand years later by Johann Gutenberg. It allowed us to expand the availability of the written word by duplicating it faster than the scribes' copying. In the 10th century, with 426 titles, the Swiss library in St. Gallen was the largest in the Christian world at the time \cite{Saleem2006}. Each of these works was a single copy accessible only to those who could afford the trip. Because of the immediate unavailability of writings, literate people were forced to memorize the works of those who preceded them (Homer, Plato, etc.); a thing which they performed with ease. 

After the introduction of the printing press, not only did libraries become repositories for countless books, but their numbers also began to increase rapidly. Before the Gutenberg machine, there were barely 30,000 books in all of Europe. Fifty years later, there were ten million printed in 236 European cities \cite{Guellec2004}. In the 10th century, there were 70 public libraries in the city of Córdoba alone \cite{Saleem2006}. 

If, as Farrukh Saleem \cite{Saleem2006} argues, whoever owns the largest library rules the world, several peoples were at a disadvantage from the outset. Today, the U.S. Library of Congress holds the largest number of volumes, with 170 million books.

\section{The Precariousness of Oral, Written, and Printed Content}
No doubt a source of pride for those who own them, libraries have nonetheless proven to be very vulnerable. Prominent libraries, including those of Alexandria (Egypt), Nalanda (India), Celsus (Italy), and Baghdad, perished in the flames, with the losses that we know. If it wasn't the libraries, it was thousands of books that were blithely burned in sadly famous book burnings\footnote{Such as those of the Spanish Inquisition in Seville in June 1481, the Nazis in Germany in May 1933, the Chinese Cultural Revolution in the 1960s, and occasionally the Communists.}. 

Thus, the survival of accumulated human knowledge was uncertain as long as it was stored on physical media such as paper or papyrus.

\section{Digitize, Digitize}
In 2004, Google decided to digitize and distribute on the Internet millions of volumes from four American libraries (Harvard, Stanford, Michigan, and the New York Public Library) and a British university (Oxford) \cite{Markoff2004}. Valued at \$150 million at the time, the project involved 15 million volumes \cite{Markoff2004}. Fifty million volumes were expected to be online by 2015. The project was a follow-up to another project, Project Gutenberg, which began in 1971 \cite{Cook2015} and had the same goal: to digitize all books in existence \cite{bookrunch2023}. 

Had these projects succeeded, the resulting libraries would invariably have been predominantly Anglo-Saxon. Already in 2011 430 languages were represented in Google Books, while there were about 7,000 in the world \cite{Lingua2022}, almost half of the titles were in English \cite{Kaplan2011}.

\section{Culture Through Datasets}
This hegemony of language persists today in the datasets on which depend Large Language Models (LLMs) such as ChatGPT, GPT 4, Perplexity, Copilot, or Gemini. A model like ChatGPT was trained on about two-thirds of the internet, the entire Wikipedia, more than 8 million documents (books, articles, websites, conversations, etc.), and more than 10 billion words \cite{CSIDDU2023}, all of which came from a technology where 55\% of the content is in English \cite{InternetSociety2023}. 

Data from developed countries are over-represented in the training datasets with which these LLMs were developed. Conversely, data from developing countries are under-represented; a very small part comes from Africa, for example. 

According to Mozilla's Internet Health Report \cite{InternetHealth2022}, from 2015 to 2020, for Africa, only Egypt's datasets were used in machine learning models. Thus, Africans in particular do not enjoy their cultures and are kept in a position of consumers of other people's data and others' data about their own cultures. 

They see themselves through the prisms of others, so to speak.

\section{The Precariousness of Digital Technology}
Unfortunately, it is not only paper and papyrus that are prone to degradation. Digital technology brings with it new challenges such as hard disk failures, computer viruses, and sloppy human manipulations. 

In 1986, the BBC Domesday incident captured people's imaginations and became a classic example of the dangers to which our digital heritage is exposed. On the occasion of the 900th anniversary of the original 1086 Domesday archive book, BBC spent £2.5 million to create a multimedia version that would fit on two laser discs. These records contained records of a million people. They also contained 50,000 photos, 3,000 datasets, 60 minutes' worth of animated images, 25,000 maps, and 250,000 place names \cite{McKie2002}. Over the years, disks had become less and less readable by computers, which were becoming more and more sophisticated. The data access problem was eventually resolved, and the images, videos, and other data could finally be visualized again. This was not done without great difficulty. Ironically, the original work was still intact after 900 years, while computer disks had not even survived for 15 years \cite{McKie2002}. 

In 1995, the U.S. government nearly lost much of its national census data due to the obsolescence of its data retrieval technology \cite{McKie2002}. In 1996, 2001, and 2002, the Internet itself came close to disaster several times. 

More recently, on March 3, 2024, a major Internet outage disrupted several social networks, particularly in countries such as India, Pakistan, and parts of East Africa \cite{Morris2024}. Most of the data traffic on the network of networks is dependent, to a very large extent, on submarine cables that are susceptible to damage by ship anchors, especially in an area as busy as the Red Sea where there are more than 15 submarine cables. On that day, four of these cables were damaged at the same time \cite{FitzGerald2024}; an exceptionally rare incident. 

Digital technologies are therefore no more immune than human memory or paper.

In October 1991, Tunisia connected to the Internet, followed by South Africa in November. The other African countries connected one by one until November 2000, when Eritrea brought up the rear.

\section{Africa?}
Africans enjoy less of the advantages of being connected to the rest of the world than of being affected by it. Culturally, today's young Africans are likely to know more about the British Empire than about the Ghanaian Empire. They are more likely to have heard of Napoleon Bonaparte than of Sundiata Keita, the founding king of the great Mali Empire. They will certainly have heard of Elon Musk as the richest man on the planet but not of Mansa Musa, king of Mali in 1312 and probably the richest man to have ever lived \cite{Evan2023,Morgan2023}. And if they have heard of Léopold Cedar Senghor or Félix Houphouët-Boigny, it will surely be thanks to the Western media. Through their prism.

\section{Digital Culture}
Unfortunately, Africa still has little data culture. Overall, Africa has the weakest statistical capacity. Indeed, statistical capacity has, over the past fifteen years, declined more in Africa than in any other region of the world \cite{Kratke2014}. Only half of African countries have conducted more than two comparable household surveys in the last ten years \cite{Tirziu2018} and only 29\% have published household surveys with education data since 2005 \cite{vanBelle2017}. 

Admittedly, African digital content already exists \cite{Gorwitz2019,Duarte2021} because African datasets do exist \cite{David2021,DeepLearningINDABA2023}. There are even African LLMs such as, for example, Kainene vos Savant, Foondamate, and MobileGPT \cite{Ogunjuyigbe2023}.

\begin{figure}[h]
\centering
\includegraphics[width=0.8\textwidth]{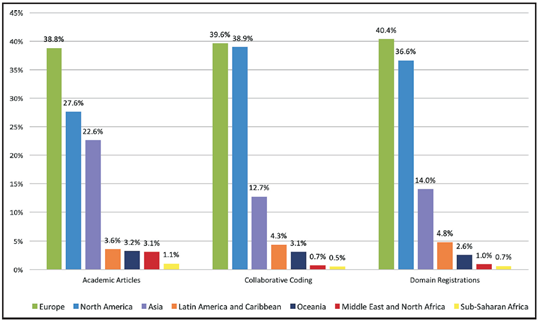}
\caption{The comparison of content creation among continents. Africa consistently lags behind other continents with percentages as low as 0.5\% \cite{Ojanpera2017}.}
\end{figure}

However, Africa continues to contribute only marginally to the global accumulation of data. According to IDC \cite{IDC2023}, each region's contributions to global data creation in 2023 were 37.4\% for North America, 32.1\% for Asia-Pacific, 19.3\% for Europe, 6.8\% for the Middle East and Africa, and 4.4\% for Latin America and the Caribbean. Sub-Saharan Africa alone is estimated to contribute 1.5\%. While Europe's contribution to Web domain registration was 40.4\%, Sub-Saharan Africa's contribution was 0.7\% \cite{Ojanpera2017} (see Figure 1). Sub-Saharan Africa contributed only 1.06\% of the global total publications in AI journals while East Asia and North America accounted for 42.87\% and 22.70\% respectively \cite{Komminoth2023}. 

While Africa (1.3 billion inhabitants) has more Internet users than North America\footnote{508.880 million compared to 444.060 according to the latest figures from Statista (2023).} (328 million inhabitants), it unfortunately has barely as many data centers as Switzerland \cite{TheEconomist2021} (8.8 million inhabitants). In 2023, the United States had 5,375 data centers \cite{Taylor2023} (2,670 in 2021 \cite{Djuraskovic2023}). Germany, second in the ranking, had 522 \cite{Taylor2023}. In the 16th place was Switzerland with 120 data centers \cite{Cloudscene2024}. 

Oftentimes even Africa's data are stored outside the continent. 

Does Farrukh Saleem's comment that whoever owns the largest library rules the world also apply to data centers and datasets?

\begin{figure}[h]
\centering
\includegraphics[width=0.8\textwidth]{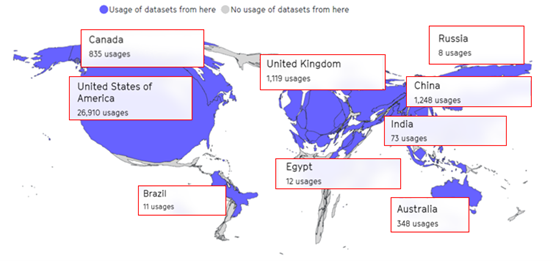}
\caption{How data sees the world. Countries are distorted based on how often the data is used in the datasets. Data usage in the U.S. accounts for the largest number of uses \cite{InternetHealth2022}.}
\end{figure}

\section{Africa's Data Presence}
While in 2016 each Internet user in the world contributed an average of 1.7 megabytes per second to the Internet \cite{FinancesOnline2024}\footnote{Some even talk about 8 megabytes per Internet user per second (Mahanti, 2022).}, African governments have begun to impose taxes on the creation of digital content of up to 15\% \cite{Agak2024,Njoya2023}. Such measures are not exactly conducive to the development of a culture of data creation. This may mean that the production of African content will cost more than the consumption of non-African content since the cost of downloading a gigabyte varies between 2\% and 16\% of monthly revenue. In Malawi, for example, a gigabyte costs an average of \$27.41 on mobile. The costs in Benin and Chad are no less exorbitant \cite{Ehl2020}. 

These policies may be contributing to the technological underdevelopment afflicting most of the continent. In sub-Saharan Africa, where 15\% of the world's population lives, only 6\% have broadband access \cite{Thompson2023}. However, there is no shortage of African initiatives. Several countries have launched digital development plans, programs, or policies. This is the case, among others, in Senegal, Rwanda, Kenya, Morocco, and South Africa. 

However, Africa faces several challenges and needs to overcome several obstacles to fully harness the potential of the data economy. 

These barriers include the increasing availability of digital devices and connectivity, the growth of the digital economy, the adoption of cutting-edge technologies like AI and the Internet of Things (IoT) as well as dealing with the growing demand for data-driven decision-making \cite{IDC2023}. For our part, we think that Africa’s inability to meet the demands of digital skills is a significant challenge we cannot overlook. Another challenge is the need for the very rapid adoption of a culture of the digital, rather than a digital culture.

\section{Conclusion}
Humanity has gone through four stages in the way it transmits knowledge: oral, written, printed, and digital. While many countries, in Africa and elsewhere, are still at the oral stage, the shift to the digital age seems inevitable if not utterly desirable. We believe that it is possible, even for countries that have not yet assimilated the stages of writing or printing, to go directly to the digital stage thanks to leapfrogging, a kind of technological shortcut. It should be remembered that digital tools do not exclude the spoken word since they are multimedia, and neither the written nor the printed are. 

With the increasing ubiquity of digital technologies in Africa, there is a pressing need to foster and promote a culture of the digital across the continent. 

This is crucial for several reasons. Firstly, promoting a culture of the digital in Africa can greatly contribute to the economic development of the continent. By embracing digital technologies, African countries can harness their potential for innovation, entrepreneurship, and job creation \cite{Friederici2020}, but a culture of the digital can enhance access to education and knowledge sharing. It can provide opportunities for individuals to gain new skills and knowledge, ultimately improving their employability and helping to bridge the educational gap in Africa \cite{Solomon2020}. 

Thus, while digital culture refers to all norms, behaviors, and practices associated with the use of digital technologies, the culture of the digital implies a focus on the cultural and social dimensions surrounding the digital itself. 

Thanks to and because of digital technologies, the influence of data on our daily lives is no longer to be demonstrated. Recent spectacular developments in AI have shown the importance of data and so-called datasets. In Africa, and in the field of AI, it is clear that more datasets will have to be created for the sake of cultural diversity and equity, though that may be more challenging at times than we think\footnote{At the end of February 2024, Google had to deal with a scandal following the images rendered by its Gemini AI. Indeed, historically and ethnically inappropriate images had been published. One example included an image of the Founding Fathers of the United States being of different races, another showed that it was a woman of color who signed the Constitution of the United States (see Kaput, 2024). Other examples include black Nazis of color, Vikings, or the Pope (see Brown (2024). In fact, Google's AI was well-intentioned, since it had tried to promote diversity in its responses, precisely to counteract the characteristic bias of these tools. The problem is obviously not limited to Google. Any LLM is fallible at this level.}. 

By the very fact that they are a reflection of the culture of their creators, the data used in the training of AIs, and particularly of LLMs, peddles biases that favor this culture and disadvantage cultures that do not carry data or lack written, printed, or digital data. 

Many African countries have launched investment policies in the digital economy, thereby boosting investment in software, ``digitization'', and ``digitalization''. Investing in and encouraging the development or, at least, the use of digital technologies are laudable policies to promote digital literacy. However, the task of building a digital culture is much more difficult because it aims to preserve and perpetuate a culture. Technology can be bought; not culture. Limiting itself to promoting the simple use of technologies invented elsewhere will relegate Africa to the role of consumer. The cornerstone of the future is the preservation of a culture, and that will only happen through data. 

Africa is renowned for its wealth of raw materials. It is time for it to invest in data, the basic raw material of a digital economy of the future in which it is the continent that will host a quarter of humanity in 2050.

\end{document}